\pdfoutput=1
\documentclass[12pt]{article}
\usepackage{graphicx}
\usepackage{lineno}
\usepackage{atlasphysics}
\usepackage{atlasbiblatex} 
\usepackage{ragged2e}
\usepackage{adjustbox}
\usepackage{wrapfig}
\usepackage{caption}

\usepackage{stackengine}
\usepackage{amsmath}

\newcommand\pubnumber{SNSN-323-63}
\newcommand\pubdate{\today}

\def\institute{Department of Physics, Royal Holloway, University of London, Surrey, TW20 0EX, UK}
\def\support{\footnote{On behalf of the ATLAS collaboration.}\footnote{Work supported by the Science and Technology Facilities Council, UK}}

\def\Title#1{\begin{center} {\Large #1 } \end{center}}
\def\Author#1{\begin{center}{ \sc #1} \end{center}}
\def\Address#1{\begin{center}{ \it #1} \end{center}}

\newcommand\pubblock{\rightline{\begin{tabular}{l} \pubnumber\\
         \pubdate  \end{tabular}}}
\newenvironment{Abstract}{\begin{quotation}  }{\end{quotation}}
\newenvironment{Presented}{\begin{quotation} \begin{center} 
             PRESENTED AT\end{center}\bigskip 
      \begin{center}\begin{large}}{\end{large}\end{center} \end{quotation}}

\def\beq{\begin{equation}}
\def\eeq#1{\label{#1}\end{equation}}
\def\eeqn{\end{equation}}

\def\beqa{\begin{eqnarray}}
\def\eeqa#1{\label{#1}\end{eqnarray}}
\def\eeqan{\end{eqnarray}}

\let\bar=\overbar

\def\Dslash{\not{\hbox{\kern-4pt $D$}}}
\def\dslash{\not{\hbox{\kern-2pt $\del$}}}

\def\msb{{\bar{\ssstyle M \kern -1pt S}}}

\begin{document}
\begin{titlepage}
\pubblock

\vfill
\Title{Test of CP Violation in $b\bar{b}$ pairs from top quark decay}
\vfill
\Author{ Jacob Kempster\support}
\Address{\institute}
\vfill
\begin{Abstract}
Top pair events provide a source of $b\bar{b}$ pairs, which can be used to probe CP violation in heavy-flavour mixing and decay. In events where one of the $W$ bosons decays leptonically to an electron or muon, the charge of the $W$ boson can be used to determine unambiguously the charge of the accompanying $b$-quark at the time of its production. In cases where the $b$-quark also decays semileptonically to a muon, this sample allows to probe two charge asymmetries constructed with the charge signs of the $W$ and the soft muon. The first measurement of the charge asymmetries in $b\bar{b}$  from top pair decays is hence presented using the data collected with the ATLAS detector during the 8 TeV run of the LHC.
\end{Abstract}
\vfill
\begin{Presented}
$9^{th}$ International Workshop on Top Quark Physics\\
Olomouc, Czech Republic,  September 19--23, 2016
\end{Presented}
\vfill
\end{titlepage}
\def\thefootnote{\fnsymbol{footnote}}
\setcounter{footnote}{0}

\section{Introduction}\label{sec:Introduction}

The violation of the combined charge conjugation (C) and parity transformation (P) of particles and antiparticles implies that the laws of physics are not the same for matter and antimatter.  However, observations of CP violation (CPV) are not sufficient to explain the matter--antimatter asymmetry in the universe. In addition, a sizeable inclusive like-sign dimuon charge asymmetry (CA) has been reported~\cite{Abazov:2013uma} by the D0 experiment in which an excess was observed over that predicted by the Standard Model (SM), when the measurement is interpreted in the form of CP asymmetries (CPAs) relevant to this study.  This is not confirmed by other experiments.

The decay products of the abundance of top quarks produced at the Large Hadron Collider (LHC)~\cite{Evans:2008zzb} makes it possible to measure CPAs in heavy-flavour mixing and decay~\cite{Gedalia:2012sx}, with a unique method of determining the charge of the $b$-quark. The top quark decays before hadronisation, predominantly via $t\rightarrow Wb$. In the case where the $W$-boson decays leptonically, the charge of the lepton determines the charge of the produced $b$-quark. The $b$-quark hadronises and in the case that the resulting $b$-hadron decays semileptonically, the charge of the soft lepton determines the $b$-quark charge at decay.  A soft-muon heavy-flavour tagging (SMT) algorithm~\cite{Aad:2015ydr} is exploited to measure such muons.

\begin{figure}[!htb]
  \centering
  \includegraphics[width=0.4\textwidth]{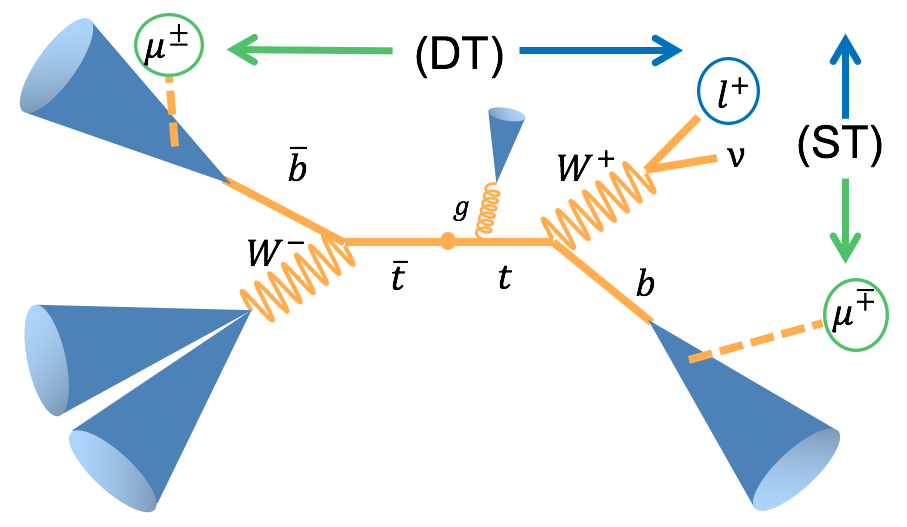}
  \caption{Illustration of ST and DT SMT muons.}
  \label{fig:AnalysisOutline:SameDiffTop}
\end{figure}

Proton--proton collisions at a centre-of-mass energy of $\sqrt{s}=8$ TeV are collected with the ATLAS~\cite{Aad:2008zzm} detector at the LHC at CERN.  Selected events have exactly one lepton and at least four jets, one of which must be tagged with the SMT algorithm and also a multivariate tagger (MV1) to increase purity.  Experimental ambiguity arises when establishing if an SMT muon originated from the same- (ST) or different-top (DT) quark as the leptonically decaying $W$-boson. This can be resolved (Section~\ref{sec:KLFitter}) and corrected for as part of the unfolding procedure. An illustration of ST and DT SMT muons is shown in Figure~\ref{fig:AnalysisOutline:SameDiffTop}. In the case of a ST SMT muon, a positively (negatively) charged $W$-boson lepton (the lepton from the decay of the $W$-boson) implies that the charge of the produced $b$-quark was negative (positive).  The opposite is true for DT SMT muons.

There are, assuming charge conjugation, three classes of ~\ttbar{} decay chains that produce two leptons of the same-sign (SS) and three which produce two leptons of opposite-sign (OS), given on the left and right of Equation~(\ref{eq:Decays}) respectively.

\noindent\begin{minipage}[t]{0.45\textwidth}
\begin{alignat}{2}
& t && \rightarrow \ell^{+}\nu\left(b\rightarrow \overline{b}\right) \rightarrow \ell^{+}\ell^{+}X, \nonumber \\
& t && \rightarrow\ell^{+}\nu\left(b\rightarrow c\right) \rightarrow \ell^{+}\ell^{+}X, \nonumber \\
& t && \rightarrow \ell^{+}\nu\left(b\rightarrow \overline{b} \rightarrow c\overline{c}\right) \rightarrow \ell^{+}\ell^{+}X, \nonumber
\end{alignat}
\end{minipage}
\noindent\begin{minipage}[t]{0.1\textwidth}
\hfill
\end{minipage}
\noindent\begin{minipage}[t]{0.45\textwidth}
\begin{alignat}{2}
& t && \rightarrow \ell^{+}\nu b \rightarrow \ell^{+}\ell^{-}X, \nonumber \\
& t && \rightarrow \ell^{+}\nu\left(b\rightarrow \overline{b} \rightarrow \overline{c}\right) \rightarrow \ell^{+}\ell^{-}X, \label{eq:Decays} \\
& t && \rightarrow \ell^{+}\nu\left(b\rightarrow  c\overline{c}\right) \rightarrow \ell^{+}\ell^{-}X. \nonumber
\end{alignat}
\end{minipage}

Observable CAs, $A^{\mathrm{ss}}$ and $A^{\mathrm{os}}$, are formed by considering the relative difference in the probability for an initial $b$- or $\overline{b}$-quark to decay via either a positively or negatively charged SMT muon. Let $N^{\alpha\beta}$ represent the number of SMT muons observed with a charge $\beta$ in conjunction with a $W$-boson lepton of charge $\alpha$, where $\alpha,\beta=\pm1$. In the case that an SMT muon is estimated to have originated from the different top-quark to the $W$-boson lepton, the sign of the $W$-boson lepton, $\alpha$, is flipped in order to consistently represent the charge of the $b$-quark at production in both scenarios.  Observable SS and OS CAs may be formed from the probabilities:

\begin{alignat}{4}
&A^{\mathrm{ss}} &&= \dfrac{P\left(b \rightarrow \ell^{+} \right) - P\left(\overline{b} \rightarrow \ell^{-}\right)}{P\left(b \rightarrow \ell^{+} \right) + P\left(\overline{b} \rightarrow \ell^{-}\right)}, \qquad &&A^{\mathrm{os}} &&= \dfrac{P\left(b \rightarrow \ell^{-}\right) - P\left(\overline{b} \rightarrow \ell^{+}\right)}{P\left(b \rightarrow \ell^{-}\right) + P\left(\overline{b} \rightarrow \ell^{+}\right)}, \label{Eq:Ass}\\[0.5ex]
&A^{\mathrm{ss}} &&= \dfrac{\left( \dfrac{N^{++}}{N^{+}} - \dfrac{N^{--}}{N^{-}} \right)}{\left( \dfrac{N^{++}}{N^{+}} + \dfrac{N^{--}}{N^{-}} \right)}, && A^{\mathrm{os}} &&= \dfrac{\left( \dfrac{N^{+-}}{N^{+}} - \dfrac{N^{-+}}{N^{-}} \right)}{\left( \dfrac{N^{+-}}{N^{+}} + \dfrac{N^{-+}}{N^{-}} \right)}. \label{Eq:Aos}
\end{alignat}

\noindent where $N^{+}\equiv N^{++}+N^{+-}$ and $N^{-}\equiv N^{-+}+N^{--}$ represent the total number of positively and negatively charged $W$-boson leptons respectively.  The CAs are expressed as ratios of probabilities to ensure that the measurements are independent of any asymmetry affecting the reconstruction of positively or negatively charged $W$-boson leptons.  The CAs are linear combinations of CPAs related to CP violation in mixing and decay: $A^{b}_{\mathrm{mix}}$, $A^{b\ell}_{\mathrm{dir}}$, $A^{bc}_{\mathrm{dir}}$ and $A^{c\ell}_{\mathrm{dir}}$.

\section{Kinematic fitting using a likelihood approach}\label{sec:KLFitter}
Separation of the data into ST- and DT-like SMT muons is performed using the kinematic likelihood fitter (KLFitter)~\cite{Erdmann:2013rxa}.  The KLFitter places Breit--Wigner mass constraints on the top-quark and $W$-boson masses, and permutes reconstructed jets into each possible position in the leading-order parton representation of the~\ttbar{} system. Transfer functions are used to map reconstructed jets to partons, and for each possible permutation a likelihood is calculated using fitted object kinematics and accounting for $b$-tagging information.  If a reconstructed $b$-tagged jet is mapped to the KLFitter leptonic (hadronic) $b$-jet position then the SMT muon is considered to be ST(DT)-like.  The KLFitter performs with a misassignment probability of $21$\%~\cite{Erdmann:2013rxa}

\section{Unfolding}\label{Unfolding}
Non-~\ttbar{} backgrounds are subtracted from the data, which are then unfolded to a well-defined particle level fiducial region (to provide a prescription with which the CPAs may be extracted from the CAs) via:

\begin{equation}
N^{i} = \frac{1}{\epsilon^{i}}\,\cdot\,\sum_{j} \mathcal{M}^{-1}_{ij}\,\cdot\,f^{j}_{\mathrm{acc}}\,\cdot\,(N^{j}_{\mathrm{data}}-N^{j}_{\mathrm{\
bkg}}),
\label{eq:fidCPV:unfoldMaster}
\end{equation}

\noindent where $i,j=\left\{{\mathrm{N^{++},N^{--},N^{+-},N^{-+}}}\right\}$ and index $i$ runs over the particle level while index $j$ runs over the reconstruction level. $N^{j}_{\mathrm{data}}$ and $N^{j}_{\mathrm{bkg}}$ are the number of SMT muons observed in data and the estimated background, respectively. $f_{\mathrm{acc}}^{j}$ and $\epsilon^{i}$ are the acceptance and efficiency terms, correcting for reconstruction (particle) level SMT muons not present at the fiducial (reconstruction) level.  $\mathcal{M}_{ij}$ is a response matrix populated exclusively by SMT muons which are matched between the reconstruction and particle level.  The CAs are then measured and the CPAs extracted, with the results presented in Table~\ref{Tab:inter:CP_comp}, alongside Mont Carlo (MC) values, SM predictions and existing experimental limits.

\begin{table}[!htb]
\begin{center}
\begin{adjustbox}{max width=\textwidth}
\begin{tabular}{l|rlrl|clrlr}
                                 & \multicolumn{2}{c}{Data $\left(10^{-2}\right)$} & \multicolumn{2}{c|}{MC $\left(10^{-2}\right)$} & \multicolumn{3}{c}{Existing limits ($2 \sigma$) $\left(10^{-2}\right)$} & \multicolumn{2}{c}{SM prediction $\left(10^{-2}\right)$} \\
\hline
$A^{\mathrm{ss}}$                 & $-$0.7 &$\pm$ 0.8  & 0.05 & $\pm$ 0.23 & \multicolumn{3}{c}{-} & $< 10^{-2}$&~\cite{Gedalia:2012sx} \\
$A^{\mathrm{os}}$                 & 0.4    &$\pm$ 0.5  & $-$0.03    & $\pm$ 0.13 & \multicolumn{3}{c}{-} & $< 10^{-2}$&~\cite{Gedalia:2012sx} \\
$A^{b}_{\mathrm{mix}}$            & $-$2.5 &$\pm$ 2.8 & 0.2    & $\pm$ 0.7 & $\quad \quad \quad$ &$< 0.1$              & ~\cite{Amhis:2014hma} & $< 10^{-3}$&~\cite{Artuso:2015swg}~\cite{Amhis:2014hma} \\
$A^{b\ell}_{\mathrm{dir}}$        & 0.5    &$\pm$ 0.5 & $-$0.03 & $\pm$ 0.14 & &$< 1.2$                & ~\cite{Descotes:2015} & $< 10^{-5}$&~\cite{Gedalia:2012sx}~\cite{Descotes:2015} \\
$A^{c\ell}_{\mathrm{dir}}$        & 1.0    &$\pm$ 1.0 & $-$0.06 & $\pm$ 0.25 & &$< 6.0$                & ~\cite{Descotes:2015} & $< 10^{-9}$&~\cite{Gedalia:2012sx}~\cite{Descotes:2015} \\
$A^{bc}_{\mathrm{dir}}$           & $-$1.0    &$\pm$ 1.1 & 0.07    & $\pm$ 0.29 & \multicolumn{3}{c}{-} & $< 10^{-7}$&~\cite{BarShalom:2010qr} \\
\end{tabular}
\end{adjustbox}
\caption{\label{Tab:inter:CP_comp} Comparison of measurements of charge asymmetries and constraints on CP asymmetries, with MC simulation (detailed in the text), existing experimental limits and SM predictions. The latter two columns represent upper limits on the absolute values $\lvert A\rvert$.}
\end{center}
\end{table}

\section{Conclusion}\label{sec:Conclusion}
A unique CPV analysis method is presented.  This analysis provides the first direct experimental limits on three direct CPAs and a limit on a mixing CPA, all consistent with the SM.



\begin{thebibliography}{99}


\bibitem{Abazov:2013uma}
D0 Collaboration, V. M. Abazov et al,
Phys. Rev. D {\bf 84} {2014} {012002}

\bibitem{Evans:2008zzb}
Evans. L et al.,
JINST {\bf 3} {2008} {S08001},

\bibitem{Gedalia:2012sx}
Gedalia. O et al.,
Phys. Rev. Lett. {\bf 110} {2013} {232002},

\bibitem{Aad:2015ydr}
ATLAS Collaboration,
JINST {\bf 11} {2016} {P04008}

\bibitem{Aad:2008zzm}
ATLAS Collaboration,
JINST {\bf 3} {2008} {S08003}

\bibitem{Erdmann:2013rxa}
Erdmann J. et al.,
NIMA {\bf 748} {2014}

\bibitem{Amhis:2014hma}
Amhis. Y et al.,
Heavy Flavor Averaging Group (HFAG) {2014}

\bibitem{Artuso:2015swg}
Artuso. M et al.,
arXiv1005.4568 {2015}

\bibitem{Descotes:2015}
Descotes-Genon S. et al.,
Phys. Rev. D {\bf 87} {2015} {074036}

\bibitem{BarShalom:2010qr}
Bar-Shalom. S et al.,
Phys. Lett. B {\bf 694} {2011}

\end{thebibliography}
\end{document}